\documentclass[fleqn,10pt]{wlscirep}
\usepackage{amsmath,amssymb}

\title{Indole moiety induced biological potency in pseudo- peptides derived from 2-amino-2-(1H-indole-2-yl) based acetamides: synthesis, structure and computational investigations}

\author[1,$\dagger$]{Kollur Shiva Prasad}
\author[2]{Renjith Raveendran Pillai}
\author[3,4,$\S$]{Madhav Prasad Ghimire}
\author[3,5,$\ddag$]{Rajyavardhan Ray}
\author[3,5]{Manuel Richter}
\author[6]{Stevan Armaković}
\author[7]{Sanja J. Armaković}
\affil[1]{Chemistry Group, Manipal Centre for Natural Sciences, Manipal Academy of Higher Education, Manipal, Udupi – 576104, Karnataka, India.}
\affil[2]{Department of Physics, TKM College of Arts and Science, Karicode, Kollam, Kerala, India.}
\affil[3]{IFW Dresden, Helmholtzstr. 20, D-01069, Dresden, Germany}
\affil[4]{Condensed Matter Physics Research Center (CMPRC), Butwal-11, Rupandehi, Lumbini, Nepal.}
\affil[5]{Dresden Center for Computational Material Science (DCMS), TU Dresden, D-01069, Dresden, Germany.}
\affil[6]{University of Novi Sad, Faculty of Sciences, Department of Physics, Trg D. Obradovića 4, 21000 Novi Sad, Serbia.}
\affil[7]{University of Novi Sad, Faculty of Sciences, Department of Chemistry, Biochemistry and Environmental Protection, Trg D. Obradovića 3, 21000 Novi Sad, Serbia.}

\affil[$\dagger$]{shivachemist@gmail.com}
\affil[$\S$]{m.p.ghimire@ifw-dresden.de}
\affil[$\ddag$]{r.ray@ifw-dresden.de}


\begin{abstract}
We report the synthesis and theoretical investigations of three novel pseudo-peptide molecules derived from 2-amino-2-(1H-indole-2-yl) acetamides. The compounds were subjected to spectroscopic characterization ($^1$H, $^{13}$C-NMR and MS) and their chemical, electronic, and optical properties have been investigated. To ascertain their potential pharmacological applicability, the prospective reactive centers and molecular sites prone to interaction with water were identified along with possible sensitivity to autoxidation. Further, we have studied the optical response in the presence of different solvents and compared the electronic and optical properties of the pristine molecules. We highlight the subtle dependence of the properties on the structure and composition of these pseudo-peptides. Our results indicate that these molecules have high pharmaceutical potential and could serve as lead components in new drug formulations.
\end{abstract}
\begin{document}

\flushbottom
\maketitle
%
%
\thispagestyle{empty}

\section*{Introduction}
Design and synthesis of peptide analogs with properties of pharmacological relevance is of topical interest.\cite{BrownNRDD2015,SoniaSciRep2016,LaxminarayanLancet2016}  This is primarily driven by two factors: (i) despite being among the most versatile bioactive molecules, peptides are known to either degrade quickly or get modified in the body,\cite{Park2003,Mahato2003} and (ii) a number of studies have shown that most of the antibacterial peptides exert their activities by enhancing the permeability of the pathogenic cell membranes. As 
a result, it is difficult to induce resistant strains against such antibacterial pathogens as compared to classic antibacterial agents.\cite{Adessi2009,White1996,Malanovic2016} In view of these, various peptidomimetic approaches are being employed to obtain peptide analogs with high functionality.\cite{Ahn2002}

	Of particular interest are pseudo-peptides or peptide bond surrogates, in which the peptide bonds are replaced with other chemical bonds.\cite{Sawyer2009} This leads to amide bond surrogates with well-defined three dimensional structures similar to natural peptides. Nevertheless, significant variation in polarity, hydrogen bonding capability and acid-base character can be achieved. Equally important is the fact that the structural and the stereo chemical integrity of the adjacent pair of $\alpha$-Carbon atoms in these pseudo-peptides remains unchanged.

	In this \emph{Report}, we present the synthesis and structural details as well as the chemical, electronic and optical properties of three newly synthesized pseudo-peptides derived from 2-amino-2-(1H-indole-2-yl), hereafter referred to as \textbf{M1}-\textbf{M3}. Detailed investigations of the reactivity, autoxidation propensity, interaction with water as a solvent, electronic and optical properties, and pharmacological potency have been carried out using Density Functional Theory (DFT) and Molecular Dynamics (MD).

	We find that the biological potency of these compounds is particularly due to indole moiety and could lead to application as an active component in some new drugs.\cite{Kaushik2013, Armakovic2012, Armakovic2015, Blessy2014, Abramovic2011, KSP2017} Specifically, there are only a few ($\leq 3$) electrophilic reaction centers and only one location particularly sensitive to autoxidation. The solvent-dependent optical properties show subtle structural and compositional effects and are understood in terms of the electronic and optical properties of the pristine molecules. Further, we show that these compounds not only obey Lipinski's rule of five\cite{Lipinski1997,Lipinski2004} but also only marginally violate the hardened Congreve rule of three conditions\cite{Congreve2003} of pharmaceutical applicability (drug likeness).

\section*{Results and Discussion}
\begin{figure}[ht!]
\centering
\includegraphics[width=0.80\linewidth]{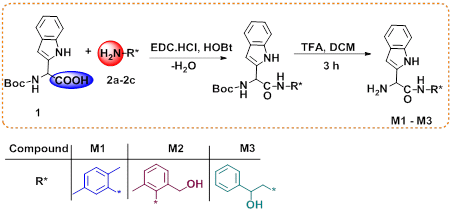}
\caption{General synthetic pathway of pseudo-peptides {\bf M1}-{\bf M3}}
\label{fig:scheme}
\end{figure}

\begin{figure}[ht!]
\centering
\includegraphics[width=0.80\linewidth]{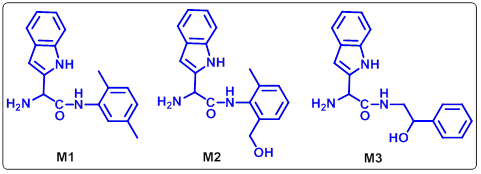}
\caption{Molecular structures of synthesized pseudo-peptides.}
\label{fig:structure}
\end{figure}

	The general synthetic route to obtain the pseudo-peptides {\bf M1}-{\bf M3} is shown in Fig. \ref{fig:scheme}. The corresponding molecular structures, shown in Fig.\ref{fig:structure}, were confirmed using spectroscopic tools, {\it viz.} NMR ($^{1}$H and $^{13}$C) and mass spectroscopy (MS) techniques. It should be noted that {\bf M1} and {\bf M2} molecules are planar while {\bf M3} is not (see Supplementary Material). This has subtle implications on the electronic properties and the charge transfer processes, as discussed in the following.

\subsection*{Reactivity, electronic properties and non-covalent interactions:}
\begin{figure}[ht!]
\centering
\includegraphics[width=0.70\linewidth]{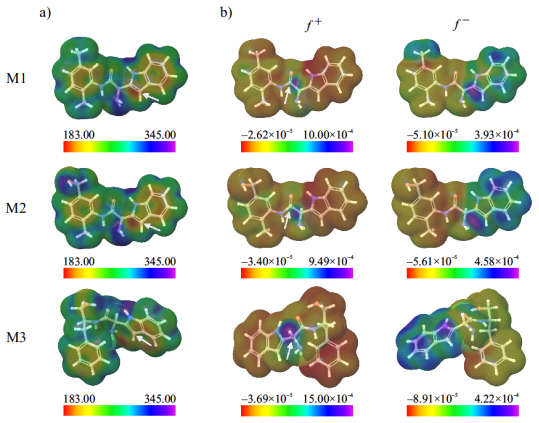}
\caption{(a) ALIE surface (left panel) and (b) $f^{+}$ and $f^{-}$ Fukui functions (center and right panels) of the {\bf M1}, {\bf M2} and {\bf M3} molecules as projected on a plane. Please note that different projections are used to highlight the results discussed in the text. The relevant C atoms are highlighted by arrows (see text for details). ALIE values are expressed in kcal/mol, while the Fukui functions are expressed in terms of electron density (electron/hohr$^3$).}
\label{fig:alie_fukui}
\end{figure}

	The molecular sites prone to electrophilic and nucleophilic additions are identified using the Average Local Ionization Energy (ALIE) surfaces and Fukui functions.\cite{Murray1990,Politzer1998,Bulat2010,Politzer2010,Mendez1994,Domingo2016} Fig. \ref{fig:alie_fukui}(a) shows the representative ALIE surfaces of the three molecules. For all the molecules, location in the near vicinity of Carbon (C) atom in the center belonging to the five membered ring (indicated by arrows) has the highest sensitivity towards electrophilic additions. These locations are characterized by the lowest ALIE values of $\sim 185$ kcal/mol.

	Significance of the central part of these molecules as reaction centers is also indicated by the Fukui functions.\cite{Mendez1994,Domingo2016} Fig. \ref{fig:alie_fukui}(b) shows the nucleophilic ($f^{+}$) and electrophilic ($f^{-}$) Fukui functions. In all the cases, the electron rich purple region for $f^{+}$ is located at the molecule’s central region. (The relevant C atoms are highlighted by arrows.) These C atoms have electrophilic nature in the case of charge addition and could be important reactive centers.
    
	On the other hand, for $f^{-}$, in all the cases, significant hole density is found over the indole part of the molecule. However, there is relatively large (hole) density at the end of the five membered ring, suggesting this site to be significant for electrophilic additions. This is in agreement with the results obtained via ALIE surfaces, as the lowest ALIE values are obtained in the vicinity of this C atom.
    
	The above features are consistent with the composition of the HOMO and LUMO states obtained via an all-electron DFT calculation of the electronic properties of the pristine molecules (see Supplementary Material). The HOMO state, which is comparable to the $f^{-}$ Fukui function, comprises of C-$p_{\rm z}$ orbitals delocalized over the five membered ring with relatively high contribution from the C atoms at the edge of the five membered ring. The LUMO state, on the other hand, is delocalized over the $R^*$ benzene ring albeit with an exceptionally high contribution from the C atom at the center of the molecule, consistent with the $f^{+}$ Fukui function discussed above. The relatively large electronegativity of the Nitrogen (N) and O atoms induces proportionate charge distributions in ground state, inducing the reactivity properties of the indole moiety.
    
	The electronic (band-) gaps are found to be approximately $2.80$ eV, $2.81$ eV and $3.37$ eV, respectively, for the {\bf M1}, {\bf M2} and {\bf M3} molecules. The significant increment in the electronic gap of the {\bf M3} molecule as compared to the others is likely due to the non-planar structure of the molecule. Indeed, such subtle implication of the structure is also found in other indicators, such as charge transfer (CT) lengths and dipole moment variations, used to study the charge transfer properties.\cite{Bahers2011}  While the charge transfer length for {\bf M3} is $\sim 40\%$ smaller than for the {\bf M1} molecule, the dipole moment change is $\sim 30\%$ larger (see Supplementary Material).
	
    Evaluation of intra-molecular non-covalent interactions (Intra-NCIs) indicate that all the molecules possess at least three such interactions between H and O atoms. Among these, the molecule {\bf M3} possesses the strongest Intra-NCI (see Supplementary Material).

\subsection*{Sensitivity towards autoxidation and water as solvent:}
	As oxidation is an important mechanism for the degradation of pharmaceutical compounds,\cite{Waterman2002, Chedaville2016,Sang2013,Kiefer2010} we study the sensitivity of these compounds to autoxidation using Hydrogen Bond Dissociation Energy (H-BDE) values.\cite{Hovorka2001,Lienard2015,Connors1986,Blanksby2003} When H-BDE values lie in the range $70-85$ kcal/mol, the sensitivity towards autoxidation of the target compound is known to be significant.\cite{Wright2009,Grynova2011,Anderson2014} Additionally, H-BDE values between $85$ and $90$ kcal/mol could also be of interest, but must be considered with caution, as these values do not necessarily mean that the molecule is sensitive towards autoxidation.\cite{Grynova2011}
    
\begin{figure}[ht!]
\centering
\includegraphics[width=0.70\linewidth]{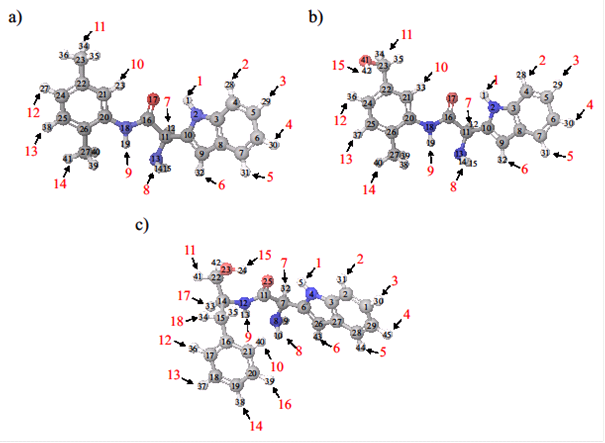}
\caption{Numeration of atoms (black) and bonds (red) for the (a) {\bf M1}, (b) {\bf M2} and (c) {\bf M3} molecules. H-BDEs have been been calculated for all inequivalent bonds, see Table \ref{table:hbde}.}
\label{fig:hbde}
\end{figure}

    The calculated values of H-BDE, listed in Table \ref{table:hbde}, provide interesting insights into the reactivity of the synthesized compounds. First, for all cases, there is one location which is highly sensitive: bond number $7$ (see Fig. \ref{fig:hbde}) for which the H-BDE value lies in the range of $75-78$ kcal/mol. This bond location is very close to the reactive center identified by the ALIE surfaces and the Fukui functions, which further emphasizes the importance of the central part of these molecules. Second, {\bf M2} and {\bf M3} have additional bonds (bonds 11 for {\bf M2}, 17 \& 18 for {\bf M3}) which could also be relevant for autoxidation. All other H-BDE values are much higher than the desired $85$ kcal/mol. The lowest H-BDE value for {\bf M3} implies that it should have the highest sensitivity towards autoxidation. Nevertheless, presence of susceptible bonds indicate significant degradation properties for all the compounds.
    
\begin{figure}[ht!]
\centering
\includegraphics[width=0.65\linewidth]{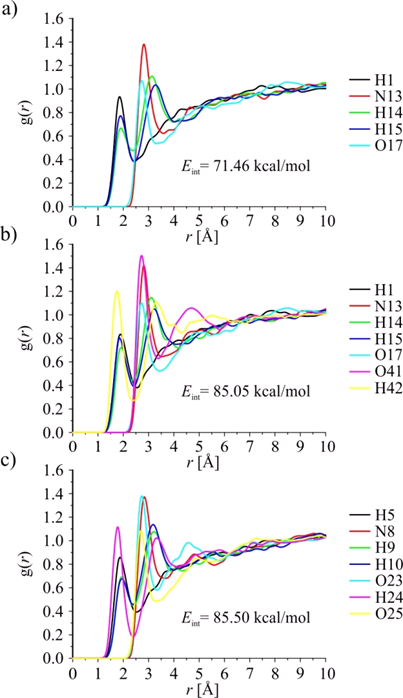}
\caption{Radial distribution functions (RDFs) of selected atoms showing significant interactions with water molecules, for (a) {\bf M1}, (b) {\bf M2} and (c) {\bf M3}.}
\label{fig:rdfs}
\end{figure}

    Since pharmaceutical molecules eventually end up in water resources, we have investigated the influence of water as a solvent in terms of interaction energy and radial distribution functions (RDFs) per atom, obtained via MD simulations. The most important RDFs and the corresponding interaction energies are presented in Fig. \ref{fig:rdfs}
All the molecules possess several H atoms that have pronounced interaction with the water molecules. While the {\bf M1} molecule has three such H atoms, {\bf M2} and {\bf M3} have four. This is also clearly reflected in the interaction energy of the molecules with water. The interaction energy of {\bf M1} is significantly lower than the corresponding values for {\bf M2} and {\bf M3}, the difference being approximately $14$ kcal/mol. The relatively stronger interactions of {\bf M2} and {\bf M3} molecules is due to the presence of OH group, which presumably leads to H-bond formation (see Fig. \ref{fig:rdfs}(b) \& (c)). The corresponding H atoms, H42 and H24 for {\bf M2} and {\bf M3}, respectively, are characterized by RDFs, $g(r)$, whose maximal values are located at distances below $2$ \AA . Other significant RDFs correspond to nitrogen and oxygen atoms, with maximal $g(r)$ values located at a distance of around $2.7$ \AA, as shown. All other RDFs have maximal $g(r)$ values at higher distances.

\subsection*{Optical properties:}
\begin{figure}[ht!]
\centering
\includegraphics[width=0.55\linewidth]{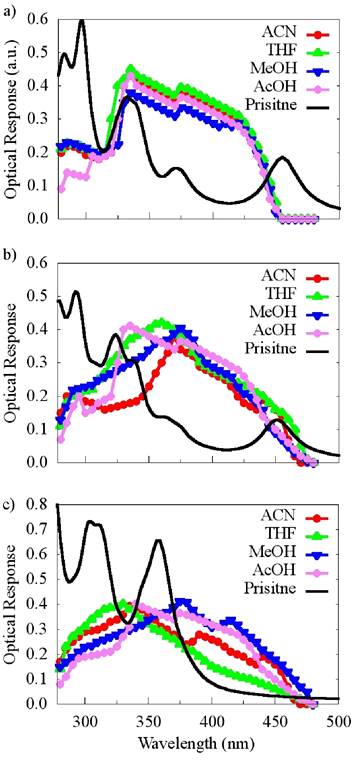}
\caption{Optical response (Absorbance) of the (a) {\bf M1}, (b) {\bf M2} and (c) {\bf M3} compounds in the presence of different solvents. A comparison is made with the corresponding theoretical calculations of the imaginary part of the dielectric function of the pristine molecules (black solid lines).}
\label{fig:optical}
\end{figure}

	 Fig \ref{fig:optical} shows the optical response (absorbance) of the molecules in the presence of different solvents. Interestingly, the absorbance curves for {\bf M1} molecule in different solvents are similar, while for the {\bf M2} \& {\bf M3} molecules, the absorbance curves differ significantly in different solvents. In order to understand these differences, we obtain the optical response of the pristine molecules in terms of the imaginary part of the dielectric constant (see Supplementary Materials). Although the two quantities are not directly comparable, it provides insights into the optical response of these systems.
	
    The low-energy optical response of the pristine {\bf M1} molecule comprises of four distinct peaks, at $455$ nm, $370$ nm, $335$ nm, and $300$ nm with the onset at approximately $475$ nm. This is in good agreement with the experimental values, suggesting that {\bf M1} is non-reactive in the presence of these solvents.
	
    As we go from {\bf M1} to {\bf M2}, the onset of optical response shifts to $455$ nm while characteristic peaks lie at $445$ nm, $377$ nm, $355$ nm, $340$ nm and $322$ nm. This is in reasonable agreement with the corresponding experimental values. It is interesting to note that presence of different solvents leads to non-uniform shifting of these peak positions. Presumably, this is due to the presence of OH group in this molecule leading to relatively high reactivity of this molecule, consistent with H-BDE values obtained earlier. In fact, such differences are further pronounced for the {\bf M3} molecule as one would expect from the H-BDE values. The characteristic feature of the optical response for the pristine {\bf M3} molecule consists of the lowest energy peak at $363$ nm, with an absorbance onset at approximately $385$ nm. The significant difference between calculated and experimental values is suggestive of the high reactivity of this molecule, consistent with the earlier results.  In addition, the differences between {\bf M3} and other molecules in the presence of solvent may also have contributions due to the non-planar structure. This feature is reflected in the direction resolved dielectric response and discussed in the Supplementary Material.
    
\subsection*{Drug likeness:}
	Table \ref{table:drug_likeness} presents a comparative analysis of the overall potential of these molecules in pharmaceutical applications (see Supplementary Material). With respect to Lipinski’s rule of five,\cite{Lipinski1997,Lipinski2004} all the molecules have fewer Hydrogen Bond Donor (HBD) and Hydrogen Bond Acceptor (HBA) sites than the upper threshold of $5$ and $10$, respectively; the number of rotatable bonds is less than the upper threshold of $10$; molecular mass is less than $500$ g/mol. Additionally, in all the cases, the molar refractivity is in the desired range, between $40$ and $130$ ${\rm m^3/mol}$.\cite{Ghose1999} Polar Surface Area (PSA) is also less than $140$ \AA$^2$. The value of one of the most important drug likeness parameters, AlogP, is less than $3$ for all the molecules, indicating outstanding permeability. It is interesting to note that while molecule {\bf M1} clearly fulfills the hardened conditions of Congreve’s rule of three,\cite{Congreve2003} {\bf M2} and {\bf M3} violate the criteria only marginally due to higher mass and number of rotatable bonds. Nevertheless, the values of the drug likeness parameters for these molecules clearly indicate high pharmaceutical potential.

\section*{Conclusions}
	In summary, the synthesis and molecular structures of three pseudo-peptide molecules are presented. Subtle differences in composition and structure have interesting consequences for the reactivity and autoxidation properties. The central part of these molecules plays a significant role in reactivity properties and autoxidation propensity inducing the indole moiety. In all cases, there are a few ($ \leq 3$) reactive centers for electrophilic additions. At the same time, there is only one site particularly sensitive to autoxidation mechanism, with very low H-BDE values of approximately $76$ kcal/mol. Presence of OH group in {\bf M2} and {\bf M3} induces larger reactivity with water, compared to {\bf M1}. Similar effects are also noted for other solvents, as obtained by comparison with the optical properties of the pristine molecules. 
	Evaluation of the drug likeness parameters indicates that all the molecules strictly satisfy the Lipinski’s rule of five. While {\bf M1} also satisfies the stringent Congreve’s rule of three, {\bf M2} and {\bf M3} violate it only marginally, suggesting that these molecules possess immense pharmaceutical potential.

\section*{Methods}


\subsection*{Synthesis, characterization and structural details}

\subsubsection*{Materials and methods}	2-amino-2-(1H-indol-2-yl) acetic acid, Di-tert-butyl dicarbonate, 2 amino-1-phenylethanol, 2-amino-3-methylbenzyl alcohol, 2-methyl-5-methylaniline and trifluoro acetic acid were purchased from Sigma-Aldrich chemical company (USA) and used as such for reactions. Solvents were purified by standard procedures and were freshly distilled prior to use. All reactions were carried out at room temperature unless mentioned otherwise.
\vspace{0.1cm}

\subsubsection*{Chemical Synthesis}

\underline{Synthesis of 2-amino-N-(2,5-dimethylphenyl)-2-(1H-indol-2-yl)acetamide (M1):}
	To a solution of Boc-protected 2-amino-1-(1H-indol-2-yl) (1) (2.5 g, 9.83 mmol) in 20 mL dichloromethane, triethyl amine (2.0 eq) was added and stirred for 20 min. The above solution was cooled (0 ºC) and then EDC.HCl (2.35 g, 12.3 mmol) was added. The mixture was stirred at this temperature for 15 min and then added HOBt (1.64 g, 12.18 mmol). After stirring for another 20 min, solution of 2-methyl-5-methylaniline (1.05 g, 9.47 mmol) in 20 mL of dichloromethane was added dropwise via syringe with continuous stirring at 0 ºC for 30 min. The resulting mixture was brought to room temperature and the stirring was continued for another 8 h under room temperature and it was monitored time to time by TLC for completion. The mixture was then quenched with ice cold water and neutralized with hydrochloric acid. The aqueous reaction mixture was then extracted with dichloromethane (twice with 25 mL portion each), and dried over anhydrous sodium sulfate. A pale yellow solid was obtained upon evaporation of solvent. The crude product obtained was the reacted with 30 \% trifluoro acetic acid in order to deprotect Boc. Finally, the product obtained was purified by column chromatography and characterized by using spectroscopic techniques. 

	Yield: 53 \%. $^{1}$H NMR (${\rm CDCl_3}$, ppm), $\delta$ 1.58 (s, 3H), 2.64 (s, 3H), 3.98 (s, 2H), 5.23 (s, 2H), 6.62 (d, 2H, $J = 8$ Hz), 7.13 (m, 7H). $^{13}$C NMR (${\rm CDCl_3}$, ppm), $\delta$ 11.52, 22.16, 60.87, 125.50, 126.10, 126.86, 127.65, 127.92, 129.27, 130.97, 134.44, 135.30, 135.82, 138.21, 167.86. ESI-MS calc. for ${\rm C_{18}H_{19}N_{3}O}$ ${\rm [M]^{+}}$: 293.1; found 293.4.

\underline{Synthesis of 2-amino-N-(5-(hydroxymethyl)-2-(1H-indol-2-yl)acetamide (M2)}
	To a solution of Boc-protected 2-amino-1-(1H-indol-2-yl) (1) (2.5 g, 9.83 mmol) in 20 mL dichloromethane was added triethyl amine (2.0 eq) and stirred for 20 min. The above solution was cooled (0 ºC) and then EDC.HCl (2.35 g, 12.3 mmol) was added. The mixture was stirred at this temperature for 15 min and then added HOBt (1.64 g, 12.18 mmol). After stirring for another 20 min, solution of 2-amino-1-phenylethane (1.11 g, 8.91 mmol) in 20 mL of dichloromethane was added dropwise via syringe with continuous stirring at 0 ºC for 30 min. The resulting mixture was brought to room temperature and the stirring was continued for another 8 h under room temperature and it was monitored time to time by TLC for completion. The mixture was then quenched with ice cold water and neutralized with hydrochloric acid. The aqueous reaction mixture was then extracted with dichloromethane (twice with 25 mL portion each), and dried over anhydrous sodium sulfate. A white solid was obtained upon evaporation of solvent. The crude product obtained was the reacted with 30 \% trifluoro acetic acid in order to deprotect Boc. Finally, the product obtained was purified by column chromatography and characterized by using spectroscopic techniques. 

	Yield: 48 \%. $^{1}$H NMR (DMSO-${\rm d_6}$, ppm), $\delta$ 2.50 (t, 1H), 4.41 (d, 3H), 5.31 (s, 2H), 5.23 (s, 2H), 6.62 (d, 2H, $J = 8$ Hz), 7.14-8.01 (m, 10H). $^{13}$C NMR (DMSO-${\rm d_6}$, ppm), $\delta$ 27.87, 31.56, 49.43, 89.78, 127.85, 127.85, 128.64, 129.06, 129.71, 130.46, 130.86, 131.49, 135.82, 171.43. ESI-MS calc. for ${\rm C_{18}H_{19}N_{3}O_{2}}$ ${\rm [M]^{+}}$: 309.15; found 310.20: ${\rm [M + H]^{+}}$.

\underline{Synthesis of 2-amino-(1-hydroxy-3-phenylpropan-2-yl)-2-(1H-indol-2yl)acetamide (M3):}
	To a solution of Boc-protected 2-amino-1-(1H-indol-2-yl) (1) (2.5 g, 9.83 mmol) in 20 mL dichloromethane was added triethyl amine (2 eq) and stirred for 20 min. The above solution was cooled (0 ºC) and then EDC.HCl (2.35 g, 12.3 mmol) was added. The mixture was stirred at this temperature for 15 min and then added HOBt (1.64 g, 12.18 mmol). After stirring for another 20 min, solution of 2-amino-3-methylbenzyl alcohol (1.18 g, 9.47 mmol) in 20 mL of dichloromethane was added dropwise via syringe with continuous stirring at 0 ºC for 30 min. The resulting mixture was brought to room temperature and the stirring was continued for another 8 h under room temperature and it was monitored time to time by TLC for completion. The mixture was then quenched with ice cold water and neutralized with hydrochloric acid. The aqueous reaction mixture was then extracted with dichloromethane (twice with 25 mL portion each), and dried over anhydrous sodium sulfate. A brownish yellow solid was obtained upon evaporation of solvent. The crude product obtained was the reacted with 30 \% trifluoro acetic acid in order to deprotect Boc. Finally, the product obtained was purified by column chromatography and characterized by using spectroscopic techniques. 
    
    Yield: 52 \%. $^{1}$H NMR (DMSO-${\rm d_6}$, ppm), $\delta$ 3.62 (d, 2H), 4.04 (d, 2H), 5.31 (s, 2H), 5.43 (s, 2H), 7.41-7.92 (m, 8H). $^{13}$C NMR (DMSO-${\rm d_6}$, ppm), $\delta$ 28.01, 45.14, 53.62, 89.55, 125.97, 127.03, 127.11, 127.61, 128.11, 128.46, 129.01, 129.16, 130.33, 135.44, 170.50. ESI-MS calc. for ${\rm C_{18}H_{19}N_{3}O_{2}}$ ${\rm [M]^{+}}$ : 309.15; found 348.20 ${\rm [M + K]^{+}}$.
	
\subsubsection*{Characterization}
	$^{1}$H and $^{13}$C-NMR spectra were obtained on Bruker Avance 400 MHz NMR spectrometer. Chemical shifts are given in ppm with respect to ${\rm SiMe_4}$ as internal standard. The deuterated solvent ${\rm CDCl_3}$ and DMSO-${\rm d_6}$ were purchased from Sigma-Aldrich chemical company. Mass spectral studies were carried out on a Q-TOF micro mass spectrometer or on a Bruker Daltonics 6000 plus mass spectrometer with ESI-MS mode analysis.

The corresponding $^{1}$H and $^{13}$C-NMR spectra and the mass spectra for the three compounds are provided in the Supplementary Material.

\subsubsection*{Structural Details}
	The structural parameters obtained from the NMR and MS spectra were used as an input to perform a complete optimization using DFT. The resulting structural parameters for the three molecules are provided in Tables S1 - S3, Supplementary Material.
    
\subsection*{Computational Details}
	\subsubsection*{Geometry optimization, ALIE, Fukui functions, and H-BDE} 
    Schr{\"o}dinger Materials Science Suite 2017-16\cite{Schrodinger} has been employed in this work in order to perform detailed comparative study. Jaguar 9.5 program\cite{jaguar} has also been used for DFT calculations in combination with Desmond\cite{desmond1,desmond2,desmond3,desmond4} and MacroModel\cite{macromodel} programs for molecular dynamics. First, a conformational search has been done with the MacroModel program to identify all possible conformations of the three synthesized molecules. Geometry optimizations with hybrid, non-local exchange and correlation functional of Becke-Lee-Yang-Parr (B3LYP)\cite{b3lyp_becke,b3lyp_lyp} has been performed with 6-31G(d) basis set in order to refine the search for the lowest energy conformation. Five lowest energy conformers of all structures have been chosen for further, more detailed, geometry optimizations with somewhat larger basis set (6-31G(d,p)) and with increased density and accuracy of grid and integrals, respectively. Frequency check for the five lowest energy conformers has also been performed in order to confirm the true ground state. Finally, the lowest energy conformer among the five lowest energy conformers in the cases of all three newly synthetized molecules have been taken for further detailed studies. The corresponding structural parameters are provided in the Supplementary Material (Tables S1 - S3). 

    ALIE surface and Fukui functions are quantum-molecular descriptors frequently employed for the visualization and determination of molecule sites prone to electrophilic additions and potential reactive centers, respectively.ALIE surface provides a (local) site dependent map of energy required for the removal of electron from the molecule. When mapped to the electron density surface, it provides information about the least tightly bound electrons. This is used to detect the molecule sites prone to electrophilic additions.

	On the other hand, the molecular electrostatic potential upon addition or removal of electron is mapped onto the local electronic density to obtain the Fukui functions. The resulting Fukui functions after the change in overall charge of the molecular systems for addition and removal of an electron are, respectively, defined as:
\begin{eqnarray}
f^{+} = \frac{\rho^{N+\delta}({\mathbf{r}}) - \rho^{N}({\mathbf{r}})}{\delta} \nonumber \\
f^{-} = \frac{\rho^{N}({\mathbf{r}}) - \rho^{N - \delta}({\mathbf{r}})}{\delta}
\end{eqnarray}
where $N$ denotes the number of electrons in reference state of the molecule, and $\delta$ is the fraction of electron which default value is set to be $0.01$.
	
     ALIE, Fukui functions and H-BDEs have been calculated with B3LYP functional and 6-311++G (d, p), 6-31+G (d, p) and 6-311G (d, p) basis sets, respectively. 
    
	\subsubsection*{MD Simulations} Simulations within MD approach have been performed with OPLS 3 force field,\cite{desmond1,opls1,opls2,opls3} where simulation time was set to 10 ns, temperature to 300 K, pressure to 1.0325 bar and cut off radius to 12 {\AA}. System was of isothermal–isobaric (NPT) ensemble class. Simple point charge (SPC) solvent model\cite{spc} was used as well. One of the corresponding molecule has been placed into the cubic box with approximately 3000 water molecules. Intramolecular noncovalent interactions were identified and visualized by electron density analysis by Johnson et al.\cite{nci1,nci2} Maestro GUI\cite{maestro} was used for the preparation of input files and analysis of results when Schrödinger Materials Science Suite 2017-1 was employed.

	\subsubsection*{Drug likeness parameters} The most important drug likeness parameter, AlogP, is defined as 1-octanol/water partition coefficient.\cite{ghose_drug1986,ghose_drug1987} Polar surface area (PSA) is defined as a surface over polar atoms. These quantities have been calculated with Maestro GUI.\cite{maestro}
    
	\subsubsection*{Electronic and Optical Properties} Further studies on electronic and optical properties were carried out for the pristine molecules using the Full-Potential Linearized Augmented Plane Wave (FP-LAPW) method of Density Functional Theory (DFT) as implemented in the WIEN2k code\cite{wien2k} 
The structural parameters obtained earlier were used for these calculations (see Tables S1 - S3, Supplementary Material).
The atomic sphere radii (RMT) were fixed at 0.66, 1.23, 1.10, 1.21 a.u. for H, N, C and O atoms, respectively. The valence states consist of 1s orbital for H, and 2s and 2p orbitals for the C, N and O atoms. The calculations were performed using the Perdew-Wang parametrization\cite{lda_pw} of the Local Density Approximation (LDA) with $R \times k_{\rm max} = 7.0$, where $k_{\rm max}$ is the plane-wave cut-off and $R$ is the smallest of all ionic radii. A $2 \times 2 \times 2$ $k$-mesh was used to obtain a self-consistent solution with vacuum of approximately 10 {\AA} in the $x$- and $y$-directions and $15$ {\AA} in the $z$-direction. The self-consistency is better than $10^{-5}$ Ry in energy and $10^{-4}$ ${\rm a.u.}^{-3}$ in charge density. In order to obtain the density of states (DOS), the linear tetrahedron method was used. 

	The optical response was studied in terms of the complex dielectric tensor, $\epsilon(\omega) =\epsilon_1(\omega) + i \epsilon_2(\omega)$, which is a measure of the linear response of the system to an external electromagnetic field. The imaginary part of the dielectric tensor is defined as\cite{optical_wien,optical_rr}:
\begin{equation}
Im \,\, \epsilon_{ij} (\omega) = \frac{4\pi}{m^2 \omega^2} \int d{\mathbf k} \sum_{nn'} 
\langle {\mathbf k}n | \vec{p}_i | {\mathbf k}n' \rangle
\langle {\mathbf k}n' | \vec{p}_j | {\mathbf k}n \rangle
\times \delta(E_{{\mathbf k}n} - E_{{\mathbf k}n'} - \hbar \omega)\,\, ,
\end{equation}
where, $i,j =$ ($x,y,z$) are the three Cartesian directions, $\vec{p}_i = \hbar \omega \nabla_i$ is the momentum operator along the direction $i$ , $|{\mathbf k}n \rangle$ is a crystal wave function with momentum ${\mathbf k}$ and band index $n$, and $\hbar \omega$ is the photon energy. The $\delta$-function is approximated as a Lorentzian with width $\Gamma =0.1$ eV for the calculations presented in this work. 

	The real part of the dielectric function and other optical properties, such as loss spectrum and refraction coefficient, can be also be obtained.\cite{optical_wien,optical_rr} The refraction coefficient is obtained from the dielectric function in the low-energy limit ($\omega \rightarrow 0$):
\begin{equation}
n(0) = \sqrt[]{\epsilon (\omega)} \,\,.
\end{equation}
For all the cases, the dielectric function was obtained by averaging the contributions along different directions while the average refractive index is calculated by taking an average of the zero-frequency limit of the real part of the refractive indices along different directions. 

	The electronic properties were further cross-checked with the Full-Potential Local-Orbital (FPLO) code\cite{fplo} (Version 14.49) using the default basis set in the 'Molecule' mode. The results obtained from the two methods were found to be in good agreement.

\bibliography{main}


\section*{Acknowledgements}
	KSP thankfully acknowledges the Director, Manipal Centre for Natural Sciences, Manipal University for financial support (start-up grant). We thank the support received from Schrödinger Inc. Part of this study was conducted within the projects supported by the Ministry of Education, Science and Technological Development of Serbia, grant numbers OI 171039 and TR 34019. MPG thanks the Alexander von Humboldt Foundation for financial support through the Georg Forster Research Fellowship Program. RR and MR acknowledge funding by the European Union (ERDF) and the Free State of Saxony via the project 100231947 (Young Investigators Group Computer Simulations for Materials Design - CoSiMa). Technical assistance from Ulrike Nitzsche was very helpful.

\section*{Author contributions statement}
K.S.P. conceived the project and carried out the synthesis. R.R.P., S.A. and S.J.A carried out part of the DFT studies (on ALIE, Fukui functions, H-BDE) and MD simulations. R.R., M.P.G. and M.R. carried out the all-electron DFT study of electronic and optical properties. All the authors contributed equally to the discussions and preparation of the manuscript.


\section*{Additional information}


The authors declare no competing financial interests.

\begin{table}[ht!]
\centering
\caption{\label{table:hbde}H-BDE values of the {\bf M1}-{\bf M3} molecules. All values are expressed in kcal/mol. For numeration of bonds, see Fig. \ref{fig:hbde}}
\begin{tabular}{|c|c|c|c|}
\hline
\hline
{\bf Bond \#} & {\bf M1} & {\bf M2} & {\bf M3} \\
\hline
\hline
1 & 100.13 & 99.69 & 98.26 \\
\hline
2 & 118.07 & 118.10 & 116.28 \\
\hline
3 & 117.64 & 117.65 & 117.52 \\
\hline
4 & 117.59 & 117.64 & 115.80 \\
\hline
5 & 117.33 & 118.24 & 117.41 \\
\hline
6 & 124.12 & 123.13 & 121.35 \\
\hline
7 & 76.99 & 77.92 & 75.27 \\
\hline
8 & 107.12 & 106.24 & 107.52 \\
\hline
9 & 105.94 & 106.48 & 113.88 \\
\hline
10 & 117.87 & 117.90 & 117.88 \\
\hline
11 & 95.10 & 87.03 & 95.50 \\
\hline
12 & 95.10 & 118.69 & 114.75 \\
\hline
13 & 116.68 & 116.64 & 115.38 \\
\hline
14 & 94.82 & 95.00 & 115.85 \\
\hline
15 & --- & 106.03 & 112.53 \\
\hline
16 & --- & --- & 115.50 \\
\hline
17 & --- & --- & 93.85 \\
\hline
18 & --- & --- & 93.06 \\
\hline
\end{tabular}
\end{table}

\begin{table}[ht]
\centering
\caption{\label{table:drug_likeness}Comparison of drug likeness parameters for the pseudo-peptide molecules {\bf M1}-{\bf M3} molecules.}
\begin{tabular}{|l|c|c|c|}
\hline
\hline
{\bf Descriptor} & {\bf M1} & {\bf M2} & {\bf M3} \\
\hline
\hline
Number of atoms & $41$ & $42$ & $45$ \\
\hline
Mass [g/mol] & $293.37$ & $309.37$ & $323.40$ \\
\hline
Number of rotable bonds & $3$ & $4$ & $6$ \\
\hline
Hydrogen bond donor (HBD) & $2$ & $3$ & $3$ \\
\hline
Hydrogen bond acceptor (HBA) & $1$ & $2$ & $2$ \\
\hline
AlogP & $2.89$ & $1.80$ & $1.73$ \\
\hline
Polar surface area (PSA) [\AA] & $70.91$ & $91.14$ & $91.14$ \\
\hline
Molar refractivity  [${\rm m^3/mol}$] & $86.95$ & $88.72$ & $92.42$ \\
\hline
\end{tabular}
\end{table}
\flushbottom

\end{document}